\documentclass[aps,groupedaddress,superscriptaddress,amsmath,amssymb,pra,twocolumn,longbibliography]{revtex4-2}
\usepackage{dcolumn}
\usepackage{bm}
\usepackage[hidelinks]{hyperref}
\usepackage{float}
\usepackage{graphicx}
\usepackage{amsfonts}
\usepackage{verbatim}
\usepackage{bbm}
\usepackage{braket}
\usepackage{qcircuit}
\usepackage[normalem]{ulem}
\usepackage[T1]{fontenc} 
\hypersetup{
    colorlinks=true,
    linkcolor=blue,
    filecolor=magenta,      
    urlcolor=cyan,
}
\usepackage{color}
\usepackage{multirow}
\usepackage{array}
\usepackage{tabularx}

\def\omq{\delta\omega_\mathrm{q}}
\def\sn{^{(n)}}

\def\ts{t_\mathrm{cyc}}

\def\h1{\mathds{1}}
\def\tT{\hat\tau}

\draft

\begin{document}
\title{Nonergodic measurements of qubit frequency noise}
\author{Filip Wudarski}
\affiliation{USRA Research Institute for Advanced Computer Science (RIACS), Mountain View, CA 94043, USA}
%
%
\author{Yaxing Zhang}
\affiliation{Google Quantum AI, Santa Barbara, CA 93111 , USA}

\author{M. I. Dykman}
\affiliation{Department of Physics and Astronomy, Michigan State University, East Lansing, MI 48824, USA}


\begin{abstract} 
Slow fluctuations of a qubit frequency are one of the major problems faced by quantum computers. To understand their origin it is necessary to go beyond the analysis of their spectra. We show that characteristic features of the fluctuations can be revealed using comparatively short sequences  of periodically repeated Ramsey measurements, with the sequence duration smaller than needed for the noise to approach the ergodic limit. The outcomes distribution and its dependence on the sequence duration are sensitive to the nature of  noise. The time needed for quantum measurements to display quasi-ergodic behavior can strongly depend on the measurement parameters.
%
\end{abstract}


\maketitle

\section{Introduction}
 \label{sec:Intro}

Due to the probabilistic nature of quantum measurements, many currently implemented quantum algorithms rely on repeatedly running a quantum computer. It is important that the qubit parameters remain essentially the same between the runs. This imposes a constraint on comparatively slow fluctuations of the qubit parameters, in particular qubit frequencies, and on developing means of revealing and charactetizing such fluctuations.   

Slow qubit frequency fluctuations have been a subject of intense studies
\cite{Nakamura2002,Bialczak2007,Alvarez2011,Bylander2011,Sank2012,Yan2012,Young2012,Paz-Silva2014,Yoshihara2014,Kim2015,Brownnutt2015,OMalley2015,Yan2016,Myers2017,Quintana2017,Paz-Silva2017,Ferrie2018,Noel2019,vonLupke2020,Proctor2020,Wolfowicz2021,Wang2021,Burgardt2023}. Of primary interest has been their spectrum, although their statistics has also attracted interest \cite{Li2013a,Ramon2015,Norris2016,Sinitsyn2016,Szankowski2017,Sung2019,Ramon2019,Sakuldee2020,Wang2020c,You2021}. This statistics is important as it may help to reveal the 
source of the underlying noise. In particular, fluctuations 
from the coupling to a few TLSs should be non-Gaussian  \cite{Paladino2002,Galperin2004,Faoro2004,Galperin2006a,Paladino2014,Ramon2015,Muller2019,Ramon2019,Huang2022}. 
The fluctuation
statistics has been often described in terms of higher-order time
correlators or their Fourier transforms, bi-spectra and high-order spectra. Most work thus far  has been done on fluctuations induced by noise 
with the correlation time smaller than the qubit decay time.

Here we show that important information about qubit frequency fluctuations 
can be extracted from what we call nonergodic measurements. The idea is to perform $M$ successive qubit measurements (for example,  Ramsey measurements) over time  longer than the qubit decay time but shorter than the noise correlation time. The measurement outcomes are determined by the instantaneous state of the 
noise 
source, for example, by the instantaneous TLSs' states.  They vary from one series of $M$ measurements to another. Thus the outcome distribution reflects the distribution of the noise source over its states. It provides information that is washed out in the ensemble averaging inherent to ergodic measurements. 

Closely related is the question of how long should a quantum measurement sequence be to reach the ergodic limit in which the noise source explores all its states. Does the measurement duration depend on the type and parameters of the measurement, not only the noise source properties, and if so, on which parameters?

A convenient and frequently used method of performing successive measurements is to repeat them periodically. In this case the  duration of data acquisition of $M$ measurements is $\propto M$. For the measurements to be nonergodic it should suffice for this duration to be smaller than the noise correlation time. This imposes a limitation on $M$ from above. The limitation on $M$ from below is imposed by the uncertainty that comes from the quantum nature of the measurements and thus requires statistical averaging over the outcomes.

\begin{figure}[h]
\includegraphics[width=0.44\textwidth]{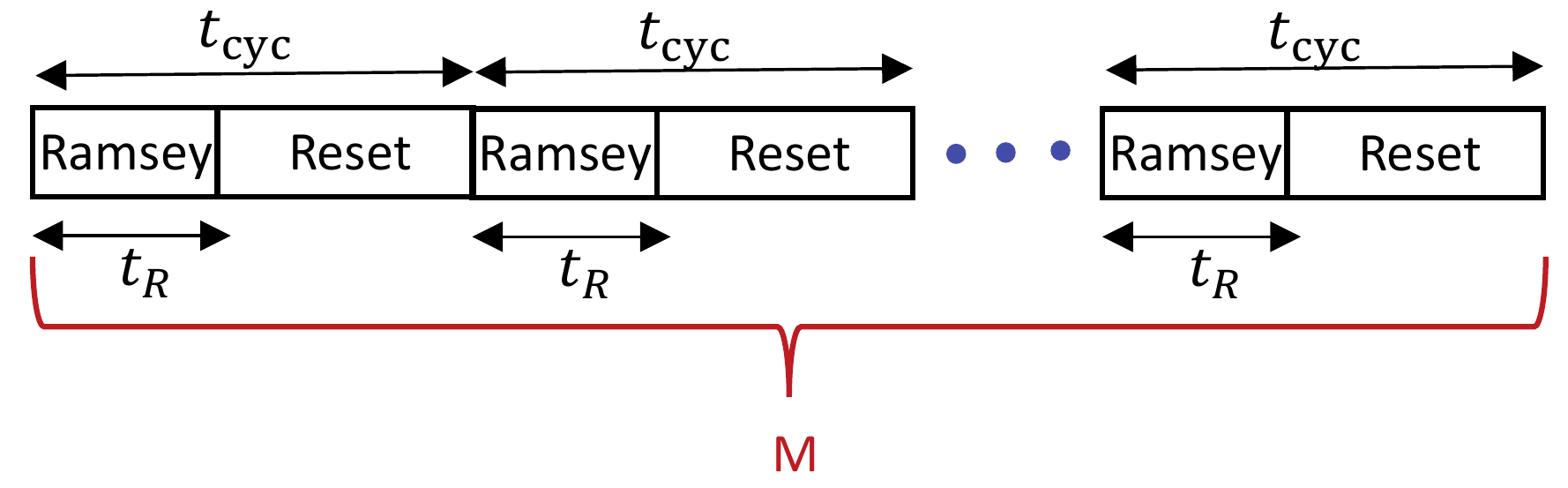} \\
\caption{Schematics of $M$ Ramsey measurements. In each measurement the qubit phase is accumulated over time $t_R$. The measurements give the probability $p$ to find the qubit in state $\ket{1}$. After a measurement the qubit is reset to state $\ket{0}$. The measurements are repeated with period $\ts$. }
\label{fig:pulse_sequence}
\end{figure}

We consider a periodic sequence of Ramsey measurements sketched in Fig.~\ref{fig:pulse_sequence}. At the beginning of a measurement the qubit, initially in the ground state $\ket{0}$,  is rotated about the $y$-axis of the Bloch sphere by $\pi/2$, which brings it  to the state $(\ket{0}+\ket{1})/\sqrt{2}$. After  time $t_R$ the rotation is repeated
and is followed by a projective measurement of finding the qubit in state $\ket{1}$. The qubit is then reset to $\ket{0}$, cf.  \cite{Fink2013}. In our scheme the measurements are repeated  $M\gg 1$ times, with period $\ts$.

The outcome of a $k$th  Ramsey measurement $x_k$ is 0 or 1. The probability $p$ to find $x_k=1$ is determined by the qubit phase accumulated over time $t_R$. This phase comes from the detuning of the qubit frequency from the frequency of the reference drive and from the noise-induced qubit frequency fluctuations $\omq(t)$.  The detuning is controllable, and we will use $\phi_R$ to indicate the phase that comes from it. The Hamiltonian $H_\mathrm{fl}$ that describes frequency fluctuations and the phase $\theta_k$ accumulated in the $k$th measurement due to these fluctuations have the form
\begin{align}
\label{eq:just_delta_omega}
H_\mathrm{fl} = -\frac{1}{2}\omq (t)\sigma_z, \qquad\theta_k = \int_{k\ts}^{k\ts+t_R}\omq (t)dt\ . 
\end{align}
Here we have set $\hbar =1$; we associate the Pauli operators $\sigma_{x,y,z}$ with the operators acting on the qubit states.  We are interested in slow frequency fluctuations. The correlation time of $\omq(t)$ is $\gtrsim\ts$ and may significantly exceed the reciprocal qubit decay rate.

In terms of the phases $\theta$ and $\phi_R$, the probability to have $x_k=1$  is \cite{Nielsen2011}
\begin{align}
\label{eq:standard_probability}
p(\theta) = \frac{1}{2}\left[1+e^{-t_R/T_2}\cos(\phi_R + \theta)\right],
\end{align}
where  $T_2^{-1}$ is the qubit decoherence  rate due to fast processes leading to decay and dephasing.
%
In the absence of qubit frequency noise $\theta=0$ for all $M$ measurements and 
the distribution of the measurement outcomes is a binomial distribution \cite{VanKampen2007}. Because of the frequency noise, the phase $\theta$ in Eq.~(\ref{eq:standard_probability}) becomes random, changing from one measurement to another, and thus the probability $p(\theta)$ also becomes random. Then the outcomes distribution  is determined not just by the quantum randomness, but also by the distribution of the values of $\theta$.  

The randomness of the phase is captured by the probability $\rho(m|M)$ to have $x_k=1$ as a measurement outcome   $m$ times in $M$ measurements, $\rho(m|M) = \mathrm{Prob}(\sum_{k=1}^M x_k=m)$.  We consider $\rho(m|M)$ for periodically repeated measurements, see Fig.~\ref{fig:pulse_sequence}. If the frequency noise has correlation time small compared to the period $\ts$, the phases $\theta_k$ in successive measurements are uncorrelated. Then $\rho(m|M)$ is still given by a binomial distribution, 
\begin{align}
\label{eq:binomial_trivial}
 \rho_\mathrm{binom}(m|M)= \binom{M}{m}r_1^m (1-r_1)^{M-m}, 
\end{align}
where  $r_1 \equiv \langle x_k\rangle =\langle p(\theta)\rangle_\theta$; here $\langle ...\rangle_\theta$ indicates averaging over realizations of $\theta$.  For large $M$ this distribution is close to a Gaussian peak centered at $r_1$.

We are interested in the opposite case of slow frequency noise. Here the distribution $\rho(m|M)$ can strongly deviate from the binomial distribution. The deviation becomes  pronounced and characteristic of the noise if $M\ts$ is comparable or smaller than the noise correlation time while $M$ is still large. 

The effect is particularly clear in the \textit{static limit}, where the noise does not change over time $M\ts$, i.e., the phase $\theta$ remains constant during $M$ measurements. Even though $\theta$ is constant, its value $\theta=\theta(\ell)$ is random, it varies from one series of $M$ measurements to another; here $\ell$ enumerates the series, and we assume that noise correlations decay between successive series. The probability $P[\theta(\ell)]$ to have a given $\theta(\ell)$ is determined by the noise statistics. The distribution of the outcomes $\rho(m|M)$ is obtained  by averaging the results of multiply repeated series of $M$ measurements. Extending the familiar arguments that lead to Eq.~(\ref{eq:binomial_trivial}), we find
\begin{align}
\label{eq:m_out_of_M_general}
&\rho(m|M)=\binom{M}{m}\sum_\ell P[\theta(\ell)] p^m[\theta(\ell)]\nonumber\\
&\times \{1-p[\theta(\ell)]\}^{M-m}
\end{align}
The distribution (\ref{eq:m_out_of_M_general}) directly reflects the distribution of the noise over its states. In particular, if the values of $\theta(\ell)$ are discrete and well-separated (see an example below), $\rho(m|M)$ has  a characteristic fine structure with peaks located at $m\approx M p[\theta(\ell))]$ for $M\gg 1$; the peak heights are determined by $P[\theta(\ell)]$.


\begin{figure*}[!tp]
      \includegraphics[width = 0.9\textwidth]{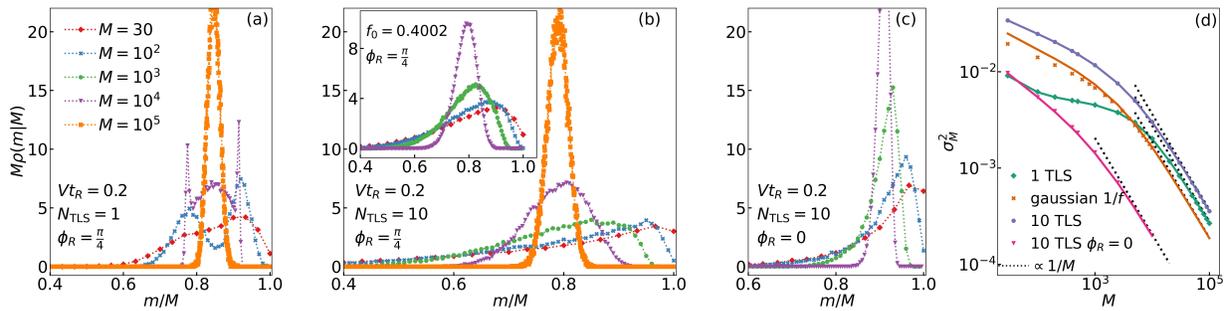}
\caption{Transition from nonerogidc to ergodic behavior with the increasing number of measurements. Red diamonds, blue crosses, green dots, purple triangles, and orange squares in panels (a)-(c) show $\rho(m|M)$ for $M=30, 10^2, 10^3, 10^4$ and $10^5$, respectively. The parameter of the dispersive qubit-to-TLS coupling is $Vt_R=0.2$. The control parameter $\phi_R$ is $\pi/4$ in panels (a) and (b).  (a) Coupling to a single symmetric TLS with the scaled switching rate $Wt_R=1.2\times 10^{-4}$. (b)  Coupling to 10 symmetric TLSs with the switching rates $W^{(n)}t_R=\exp(-3n/4 )$,  $n=3,4,\ldots,12$. The inset shows the results for Gaussian $1/f$-type noise described in the text, $f_0=\langle \delta\omega_q^2\rangle t_R^2$. (c) Coupling to the same 10 TLSs as in panel (b), but for $\phi_R=0$. (d) The variance of $\rho(m|M)$ for the data in panels (a)-(c); the solid lines show the theory, Eq.~(\ref{eq:variance_correlated_main}), the data points show the results of the simulations. The dashed lines show the ergodic limit.}
\label{fig:TLS_comparisons}
\end{figure*}

An important example of slow frequency noise is the  noise that results from dispersive coupling to a set of slowly switching TLSs, 
\begin{align}
\label{eq:q_TLS_dispersive}
\omq(t) =\sum_nV\sn{} \tT_z\sn.
\end{align}
Here $n=1,...,N_\mathrm{TLS}$ enumerates the TLSs, $\tT_z\sn$ is the Pauli operator  of the $n$th TLS, and $V\sn{} $ is the coupling parameter; the states of the $n$th TLS are $\ket{0}\sn$ and $\ket{1}\sn$,
and $\tT_z\sn\ket{i}\sn = (-1)^i\ket{i}\sn$ with $i=0,1$. We assume that the TLSs do not interact with each other. Their dynamics is described by the balance equations for the state populations. The only parameters are  the rates $W_{ij}\sn$ of $\ket{i}\sn\to \ket{j}\sn$ transitions, where $i,j=0,1$ \cite{Anderson1972,*Phillips1972,*Phillips1987}. The rates $W_{ij}\sn$ also give the stationary occupations of the TLSs states  $w_{0,1}\sn$, 
\begin{align}
\label{eq:stationary_populations}
 w_0\sn = W_{10}\sn/W\sn,\quad w_1\sn = W_{01}\sn/W\sn.
 \end{align}
Here $W\sn = W_{01}\sn + W _{10}\sn$ is the TLS relaxation rate.  The value of $\min W\sn$ gives the reciprocal correlation time of the noise from the TLSs. 

In the static-limit approximation, the TLSs remain in the initially occupied states $\ket{0}\sn$ or $\ket{1}\sn$ during all $M$ measurements. Then, from Eq.~(\ref{eq:q_TLS_dispersive}), the phase  $\delta\omega_\mathrm{q} t_R$ that the qubit  accumulates  during a  measurement is 
\begin{align}
\label{eq:phase_initial}
\theta(\{j_n\}) =  \sum_{n} V\sn{} (-1)^{j_n} t_R, 
\end{align}
where $j_n=0$ if the occupied TLS state is $\ket{0}\sn$ and $j_n=1$ if the occupied state is $\ket{1}\sn$. 
The probability to have a given $\theta(\{j_n\})$  is determined by the stationary state occupations, $P[\theta(\{j_n\})]=\prod_n w_{j_n}\sn$. 

For the TLSs' induced noise, $\ell$ in Eq.~(\ref{eq:m_out_of_M_general}) enumerates various combinations $\{j_n\}$.  With the increasing coupling  $V\sn$, the separation of the values of  $ \theta(\{j_n\})$ increases, helping to observe the fine structure of $\rho(m|M)$.

The expression for $\rho(m|M)$ simplifies in the important case where the TLSs are symmetric, $w_0\sn = w_1\sn=1/2$, and all coupling parameters are the same, $V\sn{} =V$, cf.~\cite{Ithier2005,Yoshihara2014}. In this case $\theta(\{j_n\})$ takes on values $\theta_\mathrm{sym}(\ell)=Vt_R(2\ell-N_\mathrm{TLS})$ with $0\leq \ell \leq N_\mathrm{TLS}$, and 
\begin{align}
\label{eq:m_out_of_M_TLS}
&\rho(m|M)=2^{-N_\mathrm{TLS}} \binom{M}{m}\sum_\ell  \binom{N_\mathrm{TLS}}{\ell} 
p^m[\theta_\mathrm{sym}(\ell)]
\nonumber\\  
& \times 
\{1-p[\theta_\mathrm{sym}(\ell)]\}^{M-m}.
\end{align}
The phases  $\theta_\mathrm{sym}(\ell)$ are determined by the coupling constant $V$ multiplied by the difference of the number of TLSs in the states $\ket{0}$ and $\ket{1}$, so that $\theta_\mathrm{sym}(\ell)$ may be significantly larger than for a single TLS \footnote{See Supplemental Material for more results on $\rho(m|M)$, including the fine structure, the effect of asymmetric TLSs, and the transition to the ergodic limit.}.

The probability $\rho(m|M)$ of having ``1'' $m$ times in $M$ measurements has a characteristic form also in the case of  Gaussian frequency fluctuations if the fluctuations are slow, so that $\omq(t)$ does not change over  time $M\ts$. An important example of slow noise is $1/f$ noise. In the static limit $\rho(m|M)$ is described by an extension of Eq.~(\ref{eq:m_out_of_M_general}), which takes into account that $\theta$ takes on continuous values. Respectively, one has to change in Eq.~(\ref{eq:m_out_of_M_general}) from the sum over $\ell$ to the integral over $\theta(\ell)$, with $P[\theta(\ell)]$ becoming the probability density. For Gaussian noise  $P[\theta(\ell)] = (2\pi f_0)^{-1/2}\exp[- \theta^2(\ell)/2 f_0]$, where $f_0=\langle\omq^2\rangle t_R^2$ (we assume that $\langle\omq\rangle =0$). The distribution $\rho(m|M)$ does not have fine structure, it depends only on the noise intensity in the static limit.

The opposite of the static limit is the ergodic limit, where $M\ts$ is much larger than the noise correlation time and the noise has time to explore all states during the measurements. In this limit $\rho(m|M)$ as a function of $m/M$ has a narrow peak at $r_1 =\langle m/M\rangle \equiv \sum_m(m/M)\rho(m|M)$, with $\langle [(m/M)-r_1]^n\rangle \propto M^{1-n}$.

\begin{figure}[h]
\includegraphics[width =0.4\textwidth ]{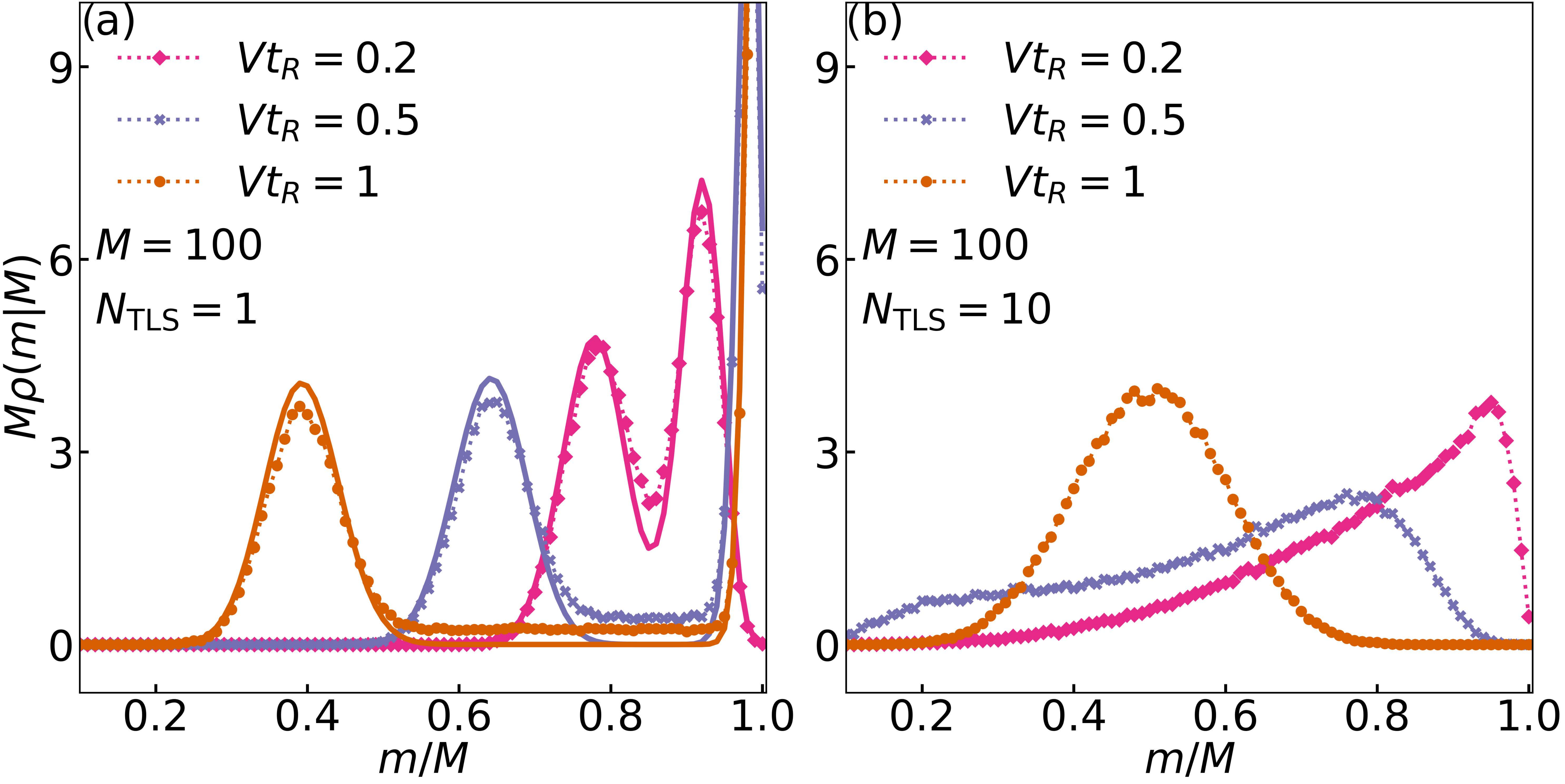}
\caption{Nonergodic behavior for a different strength  of the qubit to TLSs coupling. The results refer to $M=100$. (a) Coupling to a single TLS, $Wt_R=10^3$.  Solid lines show the static limit, Eq.~(\ref{eq:m_out_of_M_TLS}). (b) Coupling to 10 TLSs with the same switching rates as in Fig.~\ref{fig:TLS_comparisons}.
}
\label{fig:TLS_comparisons_2}
\end{figure}

{\it Simulations.}
We  performed numerical simulations to explore  the transition from the static to the ergodic  limit and the features of $\rho(m|M)$ for slow noise.  We used $\ts=3t_R$. The  measurements were simulated  at least $10^5$ times. In Figs.~\ref{fig:TLS_comparisons}  and \ref{fig:TLS_comparisons_2} we show $\rho(m|M)$ for the noise from symmetric TLSs, $W_{01}\sn = W_{10}\sn = W\sn/2$. The results for asymmetric TLSs are similar \cite{Note1}.


%
%

Figure~\ref{fig:TLS_comparisons} shows evolution of $\rho(m|M)$ with the varying measurements number $M$. It is very different for different numbers of TLSs and the measurement parameter $\phi_R$. The figure refers to a relatively weak qubit-TLS coupling. Panel (a) refers to a single TLS. Here, in the static limit $\rho(m|M)$ is double-peaked, with the peaks at $m/M\approx 0.92$ and $0.78$, from Eq.~(\ref{eq:m_out_of_M_TLS}). Such peaks are seen for $M=100$ and $M=10^4$, where $MW\ts = 3.6\times 10^{-2}$ and $3.6$, respectively, even though one might expect the system to be close to ergodic for $M=10^4$. For $M=30$ the fine structure is smeared, because $M$ is not large enough to average out the uncertainty of quantum measurements, but $\rho(m|M)$ displays a significant and characteristic asymmetry. For $M=10^5$,  where $MW\ts = 36$, the distribution does approach the ergodic limit, with a single peak at $m/M \approx 0.85$ \cite{Note1}.

Figure~\ref{fig:TLS_comparisons}~(b) refers to 10 TLSs. Their scaled switching rates $W^{(n)}\ts$ form a geometric series, varying from  $\approx 0.32$ to $\approx 3.7\times 10^{-4}$, so that the static limit does not apply and the fine structure is not resolved \cite{Note1}. The asymmetry of $\rho(m|M)$ is profound. It gradually decreases with the increasing $M$. It is important that, for $\phi_R=\pi/4$, the distribution approaches the ergodic limit for $ M\ts\times (\min W\sn)\gtrsim 30$, similar to the case of one TLS (the choice $\phi_R=\pi/4$ is explained in \cite{Note1}).  

The inset in Fig.~\ref{fig:TLS_comparisons}~(b) shows the evolution of $\rho(m|M)$ for $1/f$-type Gaussian frequency noise $\omq(t)$ with the power spectrum $S_q(\omega) = (2D/\pi)\int_{\omega_{\min}}^\infty dW/(W^2+\omega^2)$. The cutoff frequency $\omega_{\min}$ is set equal to the minimal switching rate of the 10 TLS in the main panel $\min(W\sn)$, and the intensity $D$ is chosen so that, in the ergodic limit, $\rho(m|M)$ has a maximum for the same $m/M$ as for the 10 TLSs \cite{Note1}). Yet the evolution of $\rho(m|M)$ with the increasing $M$ is fairly different from that in the main panel.

The result of Fig.~\ref{fig:TLS_comparisons}~(c) is surprising. The data refers to the same 10 TLS as in panel (b), except that the phase of the Ramsey measurement is set to $\phi_R=0$. The change of $\phi_R$ does not affect the dynamics of the TLSs. However, for the same $M$ values, the peak of $\rho(m|M)$ is much narrower than where $\phi_R=\pi/4$ and the ergodic limit is approached by the measurement outcomes for an order of magnitude smaller $M$. 

A simple measure of closeness of $\rho(m|M)$ to the ergodic limit is the variance $\sigma_M^2 = \sum_m (m/M)^2\rho(m|M)-r_1^2$, where  $r_1 = \langle x_k\rangle \equiv \sum_m(m/M)\rho(m|M)_{M\to \infty}$. A straightforward calculation shows that 
\begin{align}
\label{eq:variance_correlated_main}
    &\sigma_M^2 =M^{-1}r_1(1-r_1) +2M^{-1}\sum_{k=1}^{M-1}\tilde r_2(k)[1-(k/M)], \nonumber\\
&\tilde r_2(k) = \langle x_n x_{n+k}\rangle - \langle x_n\rangle ^2
\end{align}
($\tilde r_2(k)$ is the centered correlator of the measurement outcomes). 
For correlated noise $\rho(m|M)$ differs from the binomial distribution (\ref{eq:binomial_trivial}) and $\sigma_M^2$ is larger than its value $r_1(1-r_1)/M$ for uncorrelated noise. Still, in agreement with statistical physics, in the ergodic limit  $\sigma_M^2\propto M^{-1}$.
In contrast, the static-limit value of $\sigma_M^2 $ is generally much larger and scales differently with $M$.  

Figure~\ref{fig:TLS_comparisons}~(d) shows how $\sigma_M^2$ approaches the ergodic scaling.
For $\phi_R=\pi/4$ and the same correlation time of the noise from 1 or 10 TLSs and of Gaussian noise ($\sim 1/\min W\sn$ and $\sim 1/\omega_{\min}$), $\sigma_M^2$ behaves similarly for large $M$.
Yet, for the same 10 TLSs, but for $\phi_R=0$ the variance approaches the ergodic limit much faster.

Figure~\ref{fig:TLS_comparisons_2} shows the effect of the strength of the coupling to the TLSs for an intermediate number of measurements $M=100$. Panel~(a) shows a profoundly double-peaked distribution for a single TLS, in excellent agreement with the static-limit (\ref{eq:m_out_of_M_TLS}). As expected, the distance between the peaks increases with the increasing coupling.  For 10 TLSs, as seen  in panel~(b), the distribution is broad and strongly asymmetric. Both its shape and the position of the maximum sensitively depend on the coupling. It is important that the coupling parameter $Vt_R$ can be changed in the experiment by varying $t_R$, which helps pointing to the mechanism of the noise. We note the distinction from direct measurements of qubit frequency as a function of time \cite{Bialczak2007,Sank2012,Proctor2020}, which is efficient for still much slower noise. 

{\it Discussion.} To reach ergodic limit, a system of 10 TLSs has to visit its $2^{10}$  states. The needed time is a property of the TLSs themselves. However, the results of the measurements can approach quasi-ergodic limit, except for the far tail of the outcomes distribution, over a shorter time. This time depends on how the measurements are done. In our setup, the noise is measured by the qubit, and then the results are read through Ramsey measurements. An important parameter is the qubit-to-TLSs coupling, which we chose to be the same for all TLSs to avoid any bias. Unexpectedly,  there is another important parameter, the phase $\phi_R$.

The effect of $\phi_R$ on the convergence to the ergodic limit is not obvious in advance. It comes through the dependence on $\phi_R$ of the centered correlator $\tilde r_2(k)$ of the measurement outcomes. For   weak coupling to slowly switching TLSs, $V\sn t_R\ll 1$ and $W\sn t_R\ll 1$, and for small $\phi_R$ this correlator is small. Moreover, it falls off with the increasing $k$ much faster than for $\phi_R=\mathcal{O}(1)$ \cite{Note1}. This indicates a reduced role of the noise correlations for small $\phi_R$.  Respectively, the ergodic limit is reached must faster with the increasing $M$.

{\it Conclusions} 
We studied the distribution of the outcomes of periodically repeated Ramsey measurements with the sequence length $M\ts$ shorter than  needed to approach the ergodic limit. Such distribution proves to provide an alternative, and sensitive, means of characterizing qubit frequency noise with a long correlation time. The analytical results and simulations show that, for non-Gaussian noise, in particular the noise from TLSs, the distribution can display a characteristic fine structure. Even where there is no fine structure, the  form of the distribution and its evolution with the sequence length are noise-specific. 

The results show that the way the system approaches the ergodic limit with the increasing number of quantum measurements depends not only on the noise source, but also on the character and parameters of the measurement. These parameters are not  necessarily known in advance.  Their effect can be strong and depends on the noise source. Measurement outcomes can practically approach the ergodic limit well before the noise source approaches this limit.

FW and MID acknowledge partial support from NASA Academic Mission Services, Contract No. NNA16BD14C, and from Google under NASA-Google SAA2-403512.

 
 
%

\end{document}


\title{Supplemental materials: Nonergodic measurements of qubit frequency noise}
\author{Filip Wudarski}
\affiliation{USRA Research Institute for Advanced Computer Science (RIACS), Mountain View, CA 94043, USA}
%
%
\author{Yaxing Zhang}
\affiliation{Google Quantum AI, Santa Barbara, CA 93111, USA}

\author{M. I. Dykman}
\affiliation{Department of Physics and Astronomy, Michigan State University, East Lansing, MI 48824, USA}


\begin{abstract} 
\end{abstract}

\pacs{}

\maketitle


 
\section{Coupling to asymmetric two-level systems}
\label{sec:cplng_to_asymmetric}

Generally, the switching rates $W_{ij}$ between the states $\ket{i}$ and $\ket{j}$ of a two-level system (TLS) are different. TLSs with $W_{ij}\neq W_{ji}$ are called asymmetric. Here we present the results that show the effect of the TLS asymmetry on the qubit frequency noise for slow TLSs, where the rates $W_{ij}$ are much smaller than the reciprocal duration of the Ramsey measurement. We consider a sequence of measurements described in the main text, where Ramsey measurements are periodically repeated $M$ times with period $\ts$, see Fig. 1 of the main text. 

\begin{figure*}[!tp]
\includegraphics[width=1.0\textwidth]{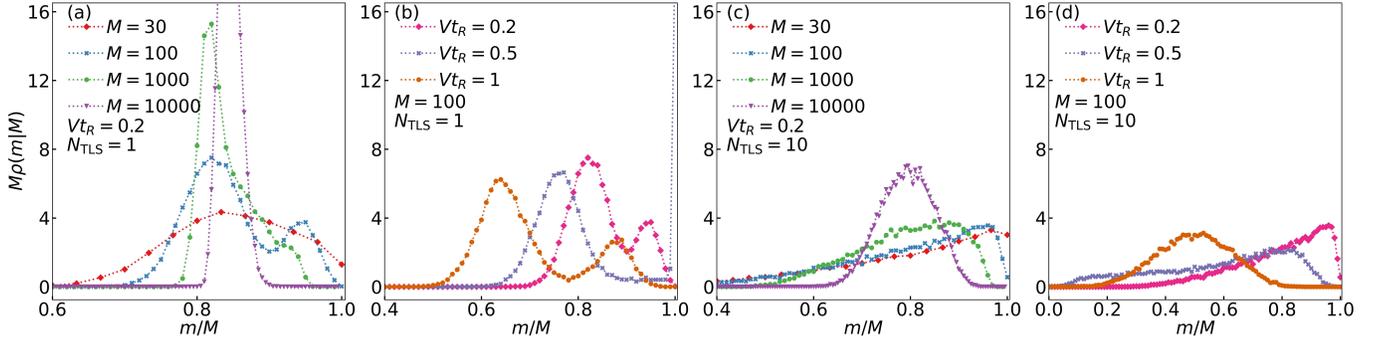}
\caption{Evolution of the probability distribution $\rho(m|M)$   for coupling to asymmetric TLSs. The measurement phase is $\phi_R=\pi/4$. The number of Ramsey measurements in a sequence $M$ and the coupling parameters are indicated in the legends. The period at which the measurements are repeated within a sequence is $\ts = 3t_R$. Panels (a) and (b) refer to one TLS with the  switching rates scaled by the duration $t_R$ of the measurement being $W_{10}t_R = 0.00075, W_{01}t_R = 0.00025$.  Panels (c) and (d) refer to 10 TLSs with the switching rates $W_{10}t_R = \exp(-3/4 n)/2, W_{01}t_R = \exp(-3/4 (n+1))/2$, where $n = 3, 4, ..., 12$. }
\label{fig:asymmetric_TLSs}
\end{figure*}

Figure~\ref{fig:asymmetric_TLSs} shows the probability distribution $\rho(m|M)$ to have $m$ ``ones'' in a sequence of $M$ Ramsey measurements of duration $t_R$, which are periodically repeated with period $\ts = 3t_R$. A characteristic feature of the distribution for the coupling to slow TLSs is the possibility to have a fine structure. The fine structure can be approximately described in the static limit, where we disregard switching between the TLS states during the time $M\ts$.  As seen from Eq. (4) of the main text, for the coupling to a single TLS with states $\ket{0}$ and $\ket{1}$, 
\begin{align}
\label{eq:fine_structure_one_TLS}
&\rho(m|M)=\binom{M}{m}\sum_{\ell=0,1} P[\theta(\ell)] p^m[\theta(\ell)]\{1-p[\theta(\ell)]\}^{M-m}, 
\nonumber\\
&\theta(\ell) = (-1)^\ell Vt_R, \qquad P[(\theta(\ell)] = W_{1-\ell\,\ell}/W
\end{align}
where $W=W_{01} + W_{10}$ is the relaxation rate of the TLS and $p(\theta) = [1+\cos(\theta+\phi_R)]/2$ is the probability of having ``one'' as a measurement outcome in the neglect of decoherence due to fast processes;  $\theta =\int_0^{t_R}dt \delta\omega_q(t)$ is the qubit phase accumulated during Ramsey measurement due to qubit frequency fluctuations $\omq(t)$; $\phi_R$ is the controlled qubit phase, see the main text.

For $M\gg 1$ the distribution (\ref{eq:fine_structure_one_TLS}) has two peaks, which are located at $m/M= p[\theta(\ell)]$  and have intensities $\propto W_{1-\ell\,\ell}/W$ ($\ell=0,1$). Such peaks are seen in panels (a) and (b) of Fig.~\ref{fig:asymmetric_TLSs} for $M=100$, where $M\ts \ll 1/W_{01}\,, 1/W_{10}$. The distance between the peaks is $\approx 2 Vt_R$. For small $M$ the structure is smeared by the uncertainty of quantum measurements. For large $M$, where $M\ts \gg 1/W_{01}\, , 1/W_{10}$, the system approaches ergodic limit, where the TLS has time to explore its states over the duration $M\ts$ of the sequence of measurement. Note a peculiar double-peak shape of the distribution for $M=10^3$ in panel (a); it refers to the parameter values where $MW_{10}\ts >1$ and $MW_{01}\ts <1$, so that the has time to switch from state $\ket{1}$, but the probability to switch from state $\ket{0}$ is still relatively small. Such structure of $\rho(m|M)$  is a feature of asymmetric TLSs.

In the case of 10 TLSs with the parameters used in Fig. ~\ref{fig:asymmetric_TLSs} the fine structure of $\rho(m|M)$ is not seen, similar to the case of symmetric TLSs presented in the main text, and for the same reason:  $M\ts W\sn >1$ for several TLSs once $M$ becomes sufficiently large, $M\gtrsim 10^2$ (we recall that $M$ has to be large to overcome the uncertainty of the outcomes of quantum measurements). The peaks of the distribution $\rho(m|M)$ are profoundly asymmetric. Their shape displays a characteristic change with the duration of the measurement sequence $M\ts$ and with the effective coupling strength controlled by the Ramsey measurement duration $t_R$.

\begin{figure}[h]
\includegraphics[width=0.48\textwidth]{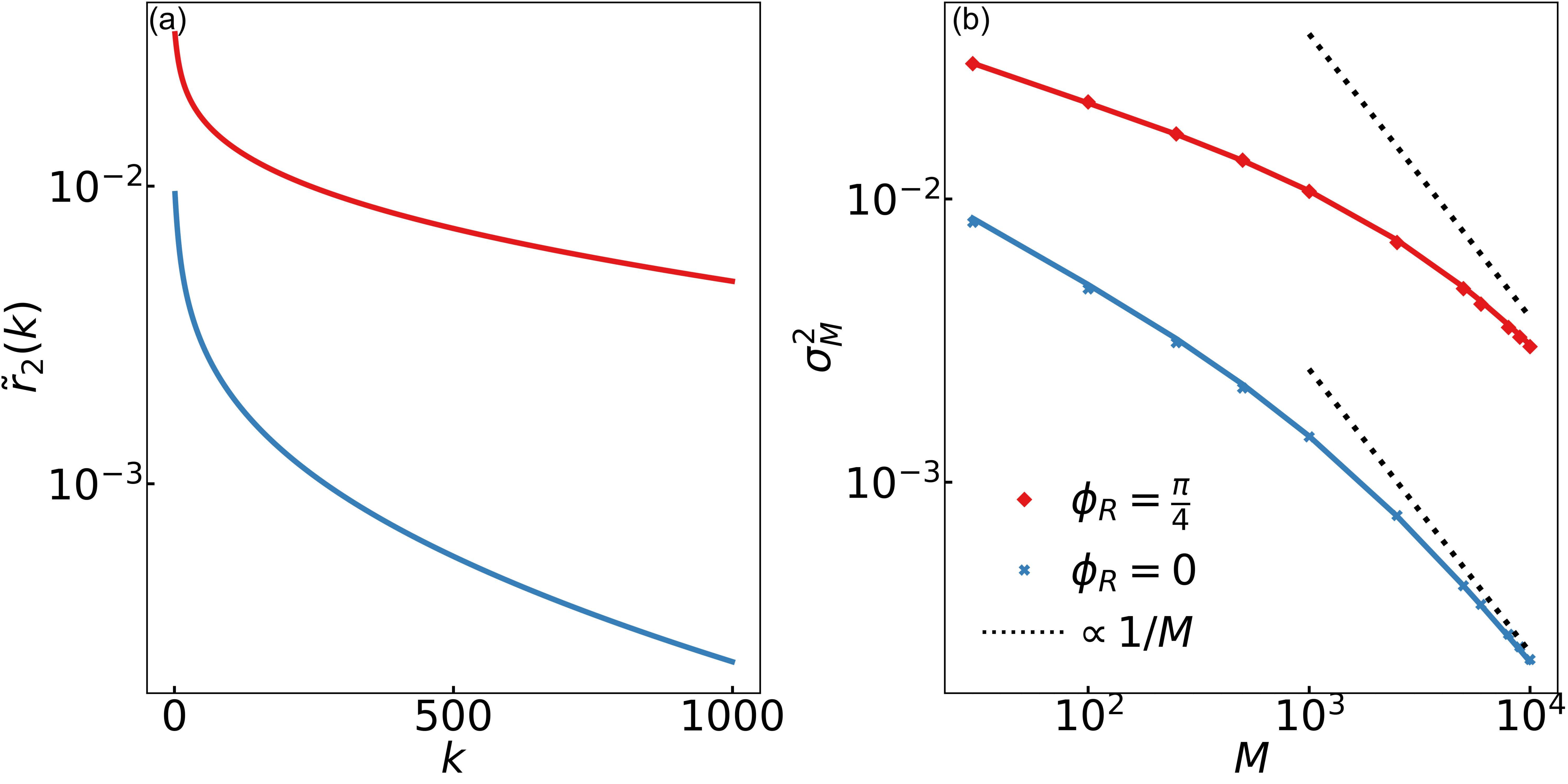}
\caption{Approaching the ergodic limit in the case of noise from asymmmetric TLSs. The figure refers to 10 TLSs with the same coupling parameter and the same switching rates as in Fig.~\ref{fig:asymmetric_TLSs}~(c). Left panel: calculated centered correlator $\tilde r_2(k)$ of the outcomes of qubit measurements, see Eqs.~(\ref{eq:correlator_defined}) and (\ref{eq:short_tR_r2}), for $\phi_R=\pi/4$ (red line) and $\phi_R=0$ (blue line). Right panel: calculated (solid lines) and simulated (dots) variance $\sigma_M^2 =  \sum_m (m/M)^2\rho(m|M)-r_1^2$ of the distribution of the measurement  outcomes. The dashed lines show the values of $\sigma_M^2$ in the ergodic limit where $\sigma_M^2\propto 1/M$. }
\label{fig:sigmaM_asymmetric}
\end{figure}

The right panel of Fig.~\ref{fig:sigmaM_asymmetric} shows how the variance of the distribution $\rho(m|M)$ approaches the ergodic limit $M\to\infty$ for the same system of 10 asymmetric TLSs as in Fig.~\ref{fig:asymmetric_TLSs} (c). As in the case of symmetric TLSs discussed in the main text and also in Sec.~\ref{sec:variance}, the variance strongly depends on the measurement phase $\phi_R$. It is much smaller for $\phi_R=0$ than for $\phi_R=\pi/4$. We explain in Sec.~\ref{sec:variance} that this is directly related to the behavior of the pair correlator $\tilde r_2(k)$, which is shown in the left panel of Fig.~\ref{fig:sigmaM_asymmetric}.


\section{Fine structure of the distribution for the coupling to multiple symmetric TLSs}
\label{sec:fine_structure_5TLSs}

\begin{figure}[h]
\includegraphics[width=0.48\textwidth]{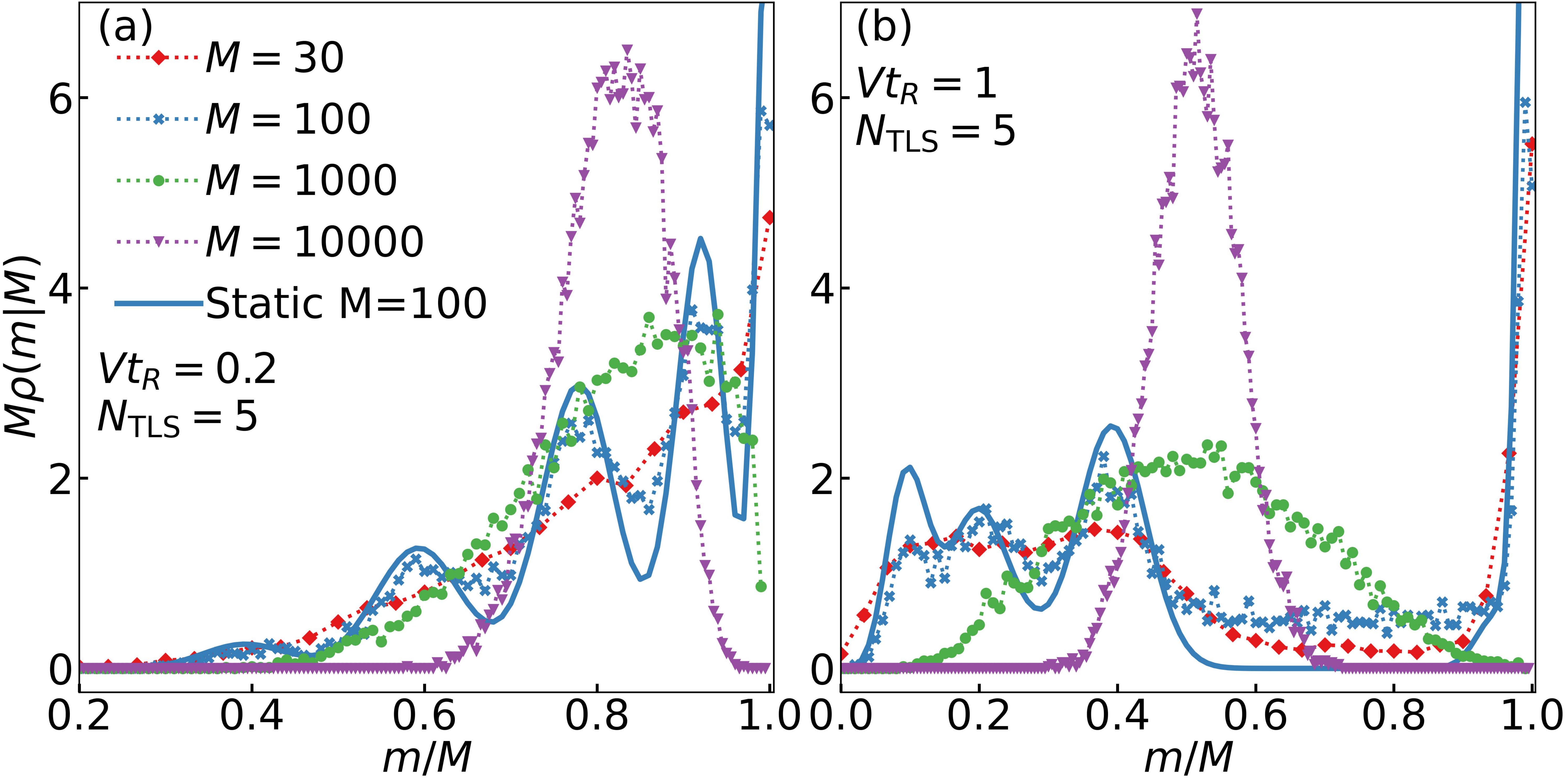}
\caption{Evolution of the probability distribution $\rho(m|M)$ with the increasing $M$ for five symmetric TLSs, $W_{ij}\sn = W_{ji}\sn = W\sn/2$. The coupling is the same for all TLSs and  $\phi_R=\pi/4$. The switching rates  are $W\sn t_R=\exp(-3n/4 )$ with $n=8, 9,\ldots,12$. The solid lines show $\rho(m|M)$ in the static limit where it is assumed that the TLSs do not switch during the measurement time $M\ts = 3Mt_R$, with $M=100$. Panels (a) and (b) refer to different coupling strengths. The color coding, which shows the values of $M$, is the same in the both panels.}
\label{fig:5_symmetric_TLSs}
\end{figure}

As indicated in the main text, the fine structure can become more pronounced for several TLSs than for one TLS, provided their coupling to the qubit is the same and their switching rates are small. Figure~\ref{fig:5_symmetric_TLSs} illustrates this effect for the coupling to five symmetric TLSs. The considered system is the five slower-switching TLSs studied in the main text, see Fig.~2 of the main text.  In the static limit the qubit phase $\theta$ can take on 6 values, $\pm Vt_R, \pm 3 Vt_R$, and $\pm 5Vt_R$, cf. Eq.~(8) of the main text. However, the theory lines in Fig.~\ref{fig:5_symmetric_TLSs} show 5 or 4 peaks, depending on the value of $Vt_R$. This is because  the probability to have ``one'' as a measurement outcome is determined by $\cos(\theta+\phi_R)$. We use $\phi_R=\pi/4$, and some peaks appear to be too small to be resolved. 

The value of $\phi_R = \pi/4$ is, in some sense, ``generic''. The measurement outcomes could be expected to be more sensitive to noise for $\phi_R=\pi/2$. In particular, observing the fine structure of the distribution $\rho(m|M)$ is easier for $\phi_R=\pi/2$, as seen from Eqs.~(4) - (8) of the main text. However, some correlators of the measurement outcomes can become zero for $\phi_R=\pi/2$, depending on the noise origin, and can be small for $\phi_R=0$ \cite{Wudarski2022}. This makes studying $\rho(m|M)$ for $\phi_R=\pi/4$ advantageous.

Overall, for $M=100$ the static-limit theory is in a reasonably good agreement with the simulations.  The numerical values of the parameter $W\sn \ts$ form a geometric series and lie between $\approx 7.4\times 10^{-3}$ and  $\approx 3.7\times 10^{-4}$. For $M=100$ the condition $M W\sn \ts \ll 1$ applies to most of the TLSs, and this is where the fine structure is most pronounced. Interestingly, for $Vt_R=1$ the fine structure is visible even for $M=30$, where the distribution $\rho(m|M)$ is broadened by the uncertainty of the measurement outcomes. For $M=10^3$ the distribution has a single broad peak for the both values of $Vt_R$. For still larger $M$ the peak narrows down, so that ultimately the distribution approaches the ergodic limit.  

\section{The centered correlator and the variance}
\label{sec:variance}

Convergence of a series of $M$ measurement outcomes to the ergodic limit depends on how long there last correlations between the outcomes. An important characteristic of the correlations is the centered pair correlator  
\begin{align}
\label{eq:correlator_defined}
\tilde r_2(k) = \Bigl\langle (x_{n+k}-\langle x_{n+k}\rangle)(x_{n}-\langle x_{n}\rangle)\Bigr\rangle
\end{align}
Here $x_k$ is the outcome of $k$th measurement, $x_k\in \{0,1\}$.  A sequence of measurements approaches the ergodic limit only when its length $M$ significantly exceeds the correlation length, i.e., $|\tilde r_2(M)/r_1|\ll 1$, where $r_1 \equiv \langle x_k\rangle$ is the average value of $x_k$. Generally, one may expect that the decay of $\tilde r_2(k)$ with the increasing $k$ is an indicator of how fast the ergodicity is approached with the increasing number of measurements $M$.

For a periodic  sequence of Ramsey measurements, the centered correlator $\tilde r_2(k)$ was calculated both for the case where the qubit frequency noise is due to the coupling to the TLSs and for the case where the noise is Gaussian \cite{Wudarski2022}. In the case of coupling to TLSs, the general expression for $\tilde r_2(k)$ simplifies in the limit where $V\sn t_R, W\sn t_R\ll 1$. One can show that, in this limit, to the leading order in the small parameters, 
\begin{align}
\label{eq:short_tR_r2}
&\tilde r_2(k) \approx \frac{1}{4}e^{-2t_R/T_2}\sin^2\phi_R
\sum_n \Bigl(w\sn V\sn{} t_R\Bigr)^2   e^{-kW\sn \ts},\nonumber\\
& w\sn = 2(W_{01}\sn W_{10}\sn)^{1/2}/W\sn,
\end{align}
 whereas 
 \begin{align}
 \label{eq:r_1_short_tR}
& r_1 \equiv \langle x_k\rangle \approx \frac{1}{2} + \frac{1}{2} e^{-t_R/T_2} \cos\phi_R\nonumber\\
 &\times \prod_n\left[1-(w\sn V\sn t_R)^2/2
 \right]; 
\end{align}
the latter expression was used in the main text to give the position of the peak of $\rho(m|M)$ in the ergodic limit.

A peculiar feature of Eq.~(\ref{eq:short_tR_r2}) is that $\tilde r_2(k)$ goes to zero for $\phi_R\to 0$. For small $\phi_R$ we need to take into account corrections of higher-order in $V\sn t_R, W\sn t_R$. Using the results \cite{Wudarski2022} one can show that, for $\phi_R=0$, 
\begin{align}
\label{eq:r2_zero_phiR}
&\tilde r_2(k) \approx \frac{1}{4}e^{-2t_R/T_2}\left\{
\sum_n \Bigl[w\sn V\sn{}^2(\Delta W\sn/W\sn) t_R^2\Bigr]^2  \right.\nonumber\\
&\left. \times  e^{-kW\sn \ts}  + \sum_{m_1=1}^{N_\mathrm{TLS}}\sum_{m_2=1}^{m_1-1}
\left(w^{(m_1)} w^{(m_2)}V^{(m_1)}V^{(m_2)}t_R^2\right)^2 \right.
\nonumber\\
&\left. \times \exp[-k(W^{(m_1)} + W^{(m_2)}) \ts] \right\}
\end{align}
where $\Delta W\sn = W_{10}\sn - W_{01}\sn$ is the difference of the switching rates of the $n$th TLS.
As seen from Eq.~(\ref{eq:r2_zero_phiR}), not only is $\tilde r_2(k)$ small for $\phi_R=0$, but also, for symmetric TLSs, it  falls off much more rapidly with the increasing $k$ than for $\phi_R=\mathcal{O}(1)$. 

\begin{figure}[h]
\includegraphics[width=0.48\textwidth]{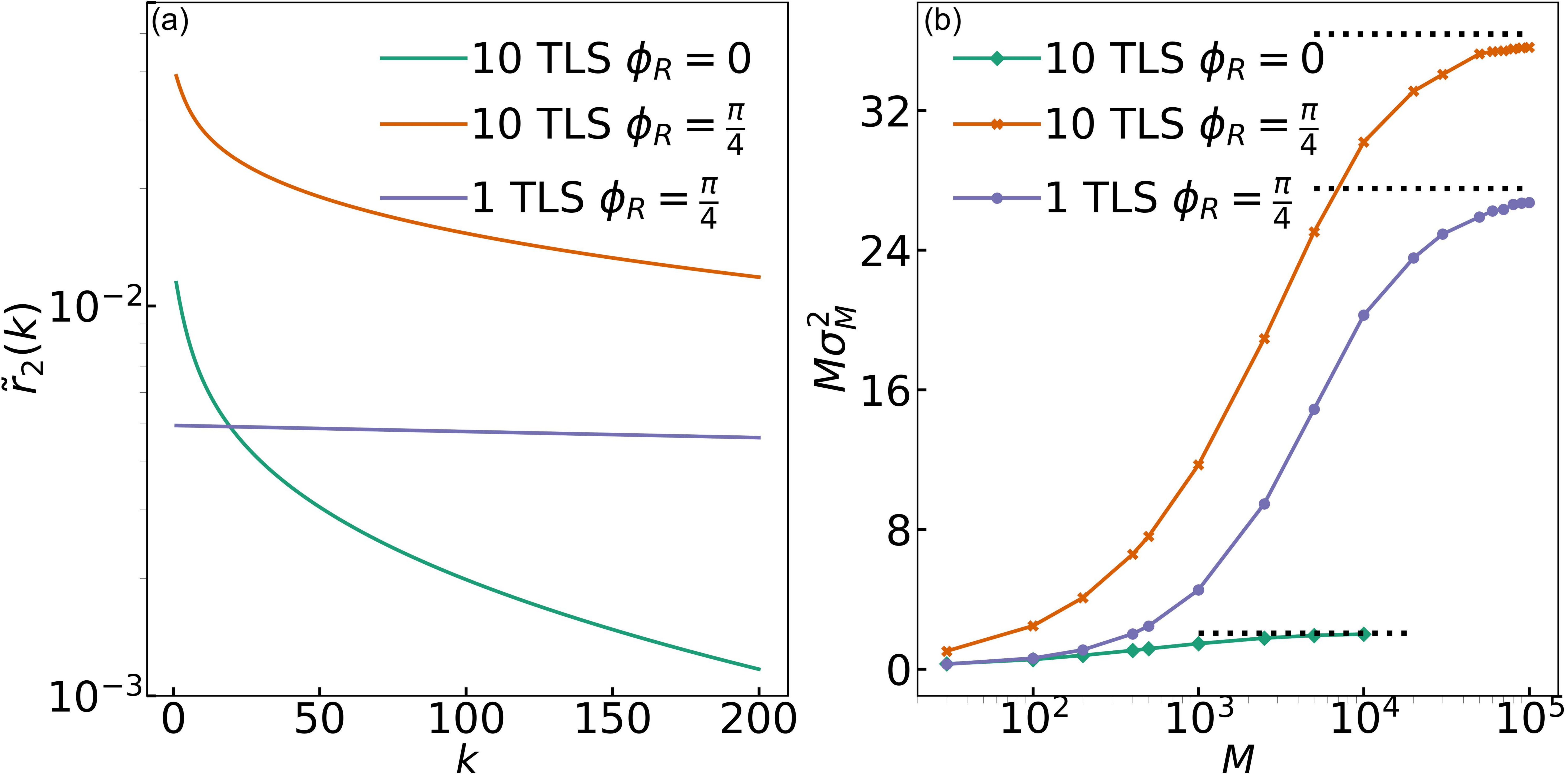}
\caption{Approaching the ergodic limit for noise from symmetric TLSs. Left panel: The calculated centered correlator $\tilde r_2(k)$ of the outcomes of qubit measurements for the quit frequency noise from TLSs.  The scaled coupling is $Vt_R=0.2$. For a single symmetric TLS the scaled switching rate $Wt_R=1.2\times 10^{-4}$. For 10 symmetric TLSs the switching rates are $W^{(n)}t_R=\exp(-3n/4 )$,  $n=3,4,\ldots,12$. 
Right panel: calculated (solid lines) and simulated (dots) variance $\sigma_M^2 =  \sum_m (m/M)^2\rho(m|M)-r_1^2$ of the distribution of the outcome measurements. The dashed lines show the values of $M\sigma_M^2$ in the ergodic limit. The TLSs parameters are the same as in the left panel.}
\label{fig:centered_correlators}
\end{figure}

Figure~\ref{fig:centered_correlators} shows  $\tilde r_2(k)$ for the coupling to symmetric TLSs for the same parameters as in Fig.~2 of the main text. As seen from the left panel, $\tilde r_2(k)$ is smaller for 1 TLS than for 10 TLSs for the same $\phi_R=\pi/4$. But the most significant difference is between the results for 10 TLSs for $\phi_R=\pi/4$ and $\phi_R=0$. This difference is responsible for the difference by an order of magnitude in the values of $M$ for which the ergodic limit is approached, as shown in the main text. It also leads to a very strong difference of the variances of the distribution  shown in the right panel. The plot of $\sigma_M^2$ here is on the linear scale, in contrast to Fig. 2 of the main text. It shows approaching the ergodic limit in more detail. 

For qubit frequency fluctuations induced by weak Gaussian noise, we have, to the leading order,
\begin{align}
\label{eq:r2_Gaussian}
&\tilde r_2(k) \approx \frac{1}{4}e^{-2(t_R/T_2)-f_0}\left (f_k\sin^2\phi_R + \frac{1}{2}f_k^2\cos^2\phi_R\right) ,\nonumber\\
&r_1 = \frac{1}{2}\left[1+ e^{-t_R/T_2}\exp(-f_0/2)\cos\phi_R\right],\nonumber\\
&f_k = \langle \theta_n\theta_{n+k}\rangle
\end{align}
(we do not expand $\exp(-f_0)$, as  $f_0$ was not very small in the simulations shown in the main text).  As seen from Eq.~(\ref{eq:r2_Gaussian}), for weak Gaussian noise $\tilde r_2(k)$ is much smaller for $\phi_R=0$, similar to the case of weak noise from the TLSs. 

In the main text, the analysis of Gaussian noise was done for $1/f$-type noise with the parameters adjusted so as to make it similar to the noise from the 10 symmetric TLSs studied in the simulations with the same coupling constants, $V\sn = V$ and $w\sn = 1/4$.. A comparison of the  noises can be done by looking at their effect on the expectation value of the measurement outcome $r_1$ and the power spectra. 

\begin{figure}[t]
\includegraphics[width=0.3\textwidth]{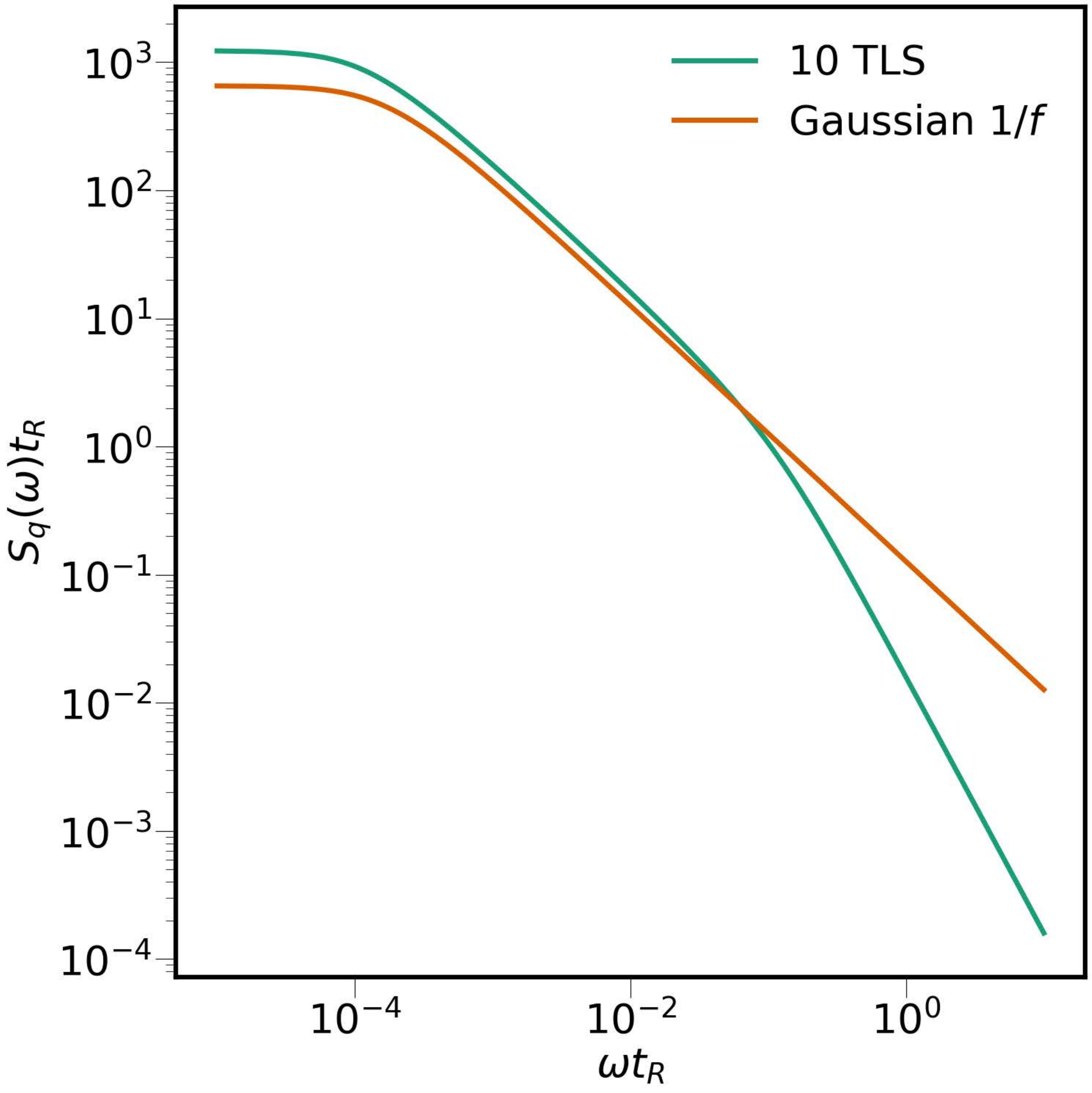}
\caption{Comparison of the power spectra of the $1/f$-type Gaussian noise and the noise from 10 symmetric TLSs. Plotted are the scaled spectra $S_q^\mathrm{TLS}(\omega)t_R$ and $S_q^\mathrm{Gauss}(\omega)t_R$. The spectra are given by Eqs.~(\ref{eq:TLS_power_spectrum}) and (\ref{eq:1_f_spectrum}). The spectra for TLSs refer to the system of 10 symmetric TLSs with the same parameters as those in Fig.~\ref{fig:centered_correlators}.  For the Gaussian noise we used $\omega_{\min} = \min W\sn$, so that $\omega_{\min} t_R=\exp(-9)\approx 1.2\times 10^{-4}$; $Dt_R = 0.1267$.}
\label{fig:power_spectra}
\end{figure}

The power spectra $S_q(\omega)$  of the noise of the qubit frequency $\omq(t)$,
\[S_q(\omega)= \int_{-\infty}^\infty dt\, e^{i\omega t}\langle \delta\omega_q(t)\delta\omega_q(0)\rangle, 
\]
are shown in Fig.~\ref{fig:power_spectra}. For the noise from  TLSs, the power spectrum has the form
\begin{align}
\label{eq:TLS_power_spectrum}
&S_q^\mathrm{TLS}(\omega)=2\sum_{n=1}^{N_\mathrm{TLS}} (w\sn V\sn)^2 W\sn 
/(W\sn{}^2+ \omega^2),
\end{align}
whereas for the Gaussian noise we studied it has the form
\begin{align}
\label{eq:1_f_spectrum}
&S_q^\mathrm{Gauss}(\omega) = \frac{2}{\pi}D_\mathrm{fl} \int_{\omega_{\min}}^\infty \frac{dW}{W^2+\omega^2}
\end{align}
As seen from Fig.~\ref{fig:power_spectra}, the spectrum $S_q^\mathrm{Gauss}(\omega)$ is of $1/f$ type, with the low-frequency offset $\omega_{\min}$. We chose $\omega_{\min}$ to be equal to the minimal switching rate of the 10 symmetric TLSs we used in the simulations, $\omega_{\min} = \min{W\sn}$. 

It is easy to see that, for $\omega_{\min} t_R\ll 1$, the parameter $f_0$ that determines $r_1$ is
\[f_0 \approx (Dt_R^2/\pi)(3/2 - \gamma_E+|\log (\omega_{\min}t_R)|.\]
The value of $f_0$ that gives the same $r_1$ as the studied 10 symmetric TLSs  was $f_0\approx 0.4$. This value was used to determine the noise intensity $D$. 

It is seen from Fig.~\ref{fig:power_spectra} that the spectra of the TLS and Gaussian noise are close at low frequencies and show $1/f$-scaling in a broad frequency range. This is why the distributions $\rho(m|M)$ for these noises in Fig. 2 of the main text were somewhat similar. Still, there was a significant differene between the distributions, which is important for identifying the source of the noise.


\section{Independent measurements}
\label{sec:independent}

It is instructive to compare the results on the distribution $\rho(m|M)$ with what is learned from independent Ramsey measurements separated by a time interval that largely exceeds the noise correlation time. Each such measurement has the probability 
\[p[\theta(\ell)]= \frac{1}{2}\left[1+\exp(-t_R/T_2)\cos\Bigl(\theta(\ell)+\phi_R\Bigr)\right]\]
to give ``one'' as an outcome. Here, the values of the phase $\theta(\ell)$ are random, they are enumerated by $\ell = 1,...,L$. 

As the measurements are repeated, we may expect to obtain a binomial distribution of the outcomes.
For the generating function of the distribution  of  the outcomes $x_k$ of independent individual measurements we have
\begin{align}
\label{eq:moments_general}
& \Big\langle \exp\left[ (t/M)\sum_{k=1}^M x_k\right]\Big\rangle =\prod_k \Big\langle \exp\left( t x_k/M\right)\Big\rangle\nonumber\\
&=\left\{1 + r_1[\exp(t/M)-1]\right\}^M, 
%
\end{align}
where $r_1\equiv \langle x_k\rangle$ is the expectation value of $x_k$.

On the other hand,  for the binomial distribution $\rho_\mathrm{binom}(m|M)$, Eq.~(3) of the main text, we have $\langle \exp(tm/M)\rangle =\{1 + r_1[\exp(t/M)-1]\}^M$.  This expression coincides with the  generating function (\ref{eq:moments_general}).



%
